\documentclass[12pt]{amsart}
\usepackage{graphicx}

\begin{document}
\centerline{\Large \bf Equilibrium distribution}
\centerline{\Large \bf on two conducting balls}
\vskip 0.5 cm
\centerline{\it Ashot Vagharshakyan}
\vskip 0.3cm
\centerline{\it Russian - Armenian
(Slavonic) University}
\centerline{\it 123 Hovsep Emin str.,
Yerevan, 0051
Armenia}
\centerline{vagharshakyan@yahoo.com}
\vskip 1cm
Keywords: {\it Conductor, equilibrium distribution,electrostatic interaction}
\vskip 0.3cm
{\it Abstract.In the paper discusses the interaction between two charged balls in equilibrium state. It is shown that, depending of the sizes, charges and distance, the balls can move in the same or opposite direction. They can repulse and attract. It is proved, that one of the balls may  vibrate.}

\section{\bf Introduction}

	S. Poisson was the first who investigate the problem of 
determining the electrostatic force between two charged balls.
	Later J. Maxwell \cite{b: M} proved, that the force of interaction between two balls with charges $Q$ and $q$, in equilibrium state, the force different from the force of interaction between point charges $Q$ and $q$ located at the centers of those balls.
	The bi-spherical coordinates \cite{b: D} are used to derive exact formulas for the force of interaction between the balls. But the results are complicated and inconvenient.	Some articles \cite{b: G}, \cite{b: J}, \cite{b: K}, \cite{b: S} are dedicated to the task of finding simpler formulas. In this paper we use the Kelvin transform and derive formulas for the equilibrium distribution of charges on each sphere.
	The obtained formulas have a relatively simple form and permit us to determine the distribution of charges on the balls' surfaces.	We also observe certain new effects: it turns out that two balls with charges of the same sign may move in the same direction and at the same time repulse and attract. In some conditions one of the balls can vibrate.

\section{\bf Location of "ghost - charges"}

	We will consider conductors of the form 
\begin{equation*}
E=B(\vec{x}_0,R)\cup B(\vec{y}_0,r),
\end{equation*}
where 
\begin{equation*}
B(\vec{a},R)=\{\vec{x}:\,\, \|\vec{x}-\vec{a}\|\leq R\}.
\end{equation*} 
Assume that the ball $B(\vec{x}_0,R)$ carry the charge $Q$ 
and the ball $B(\vec{y}_0,r)$ carry the charge $q$. Denote by
\begin{equation*}
d=\|\vec{x}_0-\vec{y}_0\|>R+r.
\end{equation*}
the distance between the centers of the balls. 

Let us define $\vec {x}_1$ by the formula 
\begin{equation*}
\vec{x}_1=\vec{x}_0+(\vec{y}_0-\vec{x}_0)\frac{R^2}{\| \vec{y}_0-\vec{x}_0\|^2}.
\end{equation*}
The point $\vec {x}_1$ is symmetric to $\vec{y}_0$ with respect to the sphere 
$\partial B(\vec x_0,R)$. 
Similarly, the point 
\begin{equation*}
\vec{y}_1=\vec{y}_0+(\vec{x}_0-\vec{y}_0)\frac{r^2}{\| \vec{x}_0-\vec{y}_0\|^2}
\end{equation*}
is symmetric to $\vec{x}_0$ with respect to the sphere $\partial B(\vec y_0,r)$.

The points $\vec{x}_{n}$ and $\vec{y}_{n}$ are the location of "ghost - charges". Note, that 
\begin{equation*}
\vec{x}_n\in B(\vec{x}_0,R),\quad \vec{y}_n \in B(\vec{y}_0,r),\quad n=0,1,2,\dots .
\end{equation*}
	At the points $\vec{x}_n,\,\,\vec{y}_n$ let us put charges $Q_n,\,\,q_n$. The "ghost - charges" itself do not appear in equilibrium distribution, but out of conductor $E$, generate the same potential function as equilibrium distribution.

Since the points $\vec{y}_{m-1},\,\vec{x}_{m},\,\,n=1,2,\dots$ are symmetric with respect to the sphere $\partial B(\vec{x}_0,R)$ so
\begin{equation*}
\frac{1}{\|\vec{x}-\vec{y}_{m-1}\|}=
\frac{\|\vec{y}_0-\vec{y}_{m-1}\|}{R}
\frac{1}{\|\vec{x}_{m}-\vec{x}\|},\quad \vec{x}\in \partial B(\vec{x}_0,R).
\end{equation*}

	Let us introduce two sequences of nonnegative numbers $u(n)$ and $v(n)$ by the formulas
\begin{equation*}
u(n)=\|\vec{x}_n-\vec{x}_0\|,\quad 
v(n)=\|\vec{y}_n-\vec{y}_0\|,\quad n=0,1,2,\dots
\end{equation*}
We have $u(0)=v(0)=0$ and
\begin{equation*}
u(n)=\frac{r^2}{d-v(n-1)},\quad 
v(n)=\frac{R^2}{d-u(n-1)},\quad n=1,2,\dots.
\end{equation*}
	Let us note that the following inequalities hold:
\begin{equation*}
u(n)\leq \frac{R^2}{d-r}<R,\quad 
v(n)\leq \frac{r^2}{d-R}<r, \quad n=1,2,\dots,
\end{equation*}

\section{\bf The "ghost - charges"}

The "ghost - charges" are
\begin{equation*}
\sum_{n=0}^{\infty} Q_{n}\delta(\vec{x}-\vec{x}_n)+ 
\sum_{n=0}^{\infty}q_{n}\delta(\vec{x}-\vec{y}_n)
\end{equation*}
	For an arbitrary numbers $Q_0,\,\,q_0$ define the sequences $Q_n,\,\,q_n\colon \quad n=1,2,\dots$ by the following recursive formulas
\begin{equation*}
Q_{n}=-\frac{u(n)}{R} q_{n-1},\quad 
q_{n}=-\frac{v(n)}{r} Q_{n-1},\quad n=1,2,\dots .
\end{equation*}
	Let for some $M<\infty$ the inequalities $|Q_0|,\,\,|q_0|\leq M$ hold. Then
\begin{equation*}
|q_{2m}|,\,\,|Q_{2m}|\leq 
M\left(\frac{Rr}{(d-R)(d-r)}\right)^m,
\quad m=1,2,3,\dots .
\end{equation*}
These inequalities we can write in the form 
\[
|Q_n|^2,\,\,|q_n|^2\leq M^2\left(\frac{Rr}{(d-R)(d-r)}\right)^n,
\quad n=0,1,2,\dots .
\]
Since $ Rr<(d-R)(d-r)$ so the sequences $Q_n,\,\,q_n$ 
tend to zero as geomteric progression. 

\section{\bf The Potential Function of the equilibrium distribution}

Let us denote
\begin{equation}
U\left(\vec{x}\right)=\sum_{n=0}^{\infty}
\frac{Q_n}{\|\vec{x}-\vec{x}_n\|}+\sum_{n=0}^{\infty}
\frac{q_n}{\|\vec{x}-\vec{y}_n\|},\,\,\,\vec{x}\notin E.
\end{equation}
The total charge $Q$ of the ball $B(\vec{x}_0,R)$ equals:
\begin{equation*}
Q=\sum_{n=0}^{\infty}Q_n=Q_0\left(1+\frac{u(2)v(1)}{rR}+\dots\right)
-q_0\left(\frac{u(1)}{R}+\frac{u(1)u(3)v(2)}{R^2r}\dots\right).
\end{equation*}
Similarly, the total charge $q$ of the ball $B(\vec{y}_0,r)$ equals:
\begin{equation*}
q=\sum_{n=0}^{\infty}q_n=
q_0\left(1+\frac{u(1)v(2)}{rR}+\dots\right)-
Q_0\left(\frac{v(1)}{r}+\frac{u(2)v(1)v(3)}{Rr^2}\dots\right).
\end{equation*}
	It is well-known that the potential function of the equilibrium distribution $U(\vec{x})$ uniquely  determined by the following properties

1.
\begin{equation*}
\Delta U(\vec{x})=0\quad \vec{x}\notin E;
\end{equation*}

2. 
\begin{equation*}
U(\vec{x})=const ,\,\, \vec{x}\in B(\vec{x}_0,R);
\end{equation*}

3. 
\begin{equation*}
U(\vec{x})=const ,\,\, \vec{x}\in B(\vec{y}_0,r);
\end{equation*}

4.
\begin{equation*}
\lim_{\|\vec{x}\|\to \infty}U(\vec{x})=0.
\end{equation*}

It is obvious that the function (1) satisfies conditions 1, and 4.

We have 
\begin{equation*}
U\left(\vec{x}\right)=
\frac{Q_0}{\|\vec{x}-\vec{x}_0\|}+q_0\left(\frac{1}{\|\vec{x}-\vec{y}_0\|}-
\frac{\|\vec{x}_1-\vec{x}_0\|}{R\|\vec{x}-\vec{x}_1\|}\right)+
\end{equation*}
\begin{equation*}
+\sum_{n=1}^{\infty}q_n\left(\frac{1}{\|\vec{x}-\vec{y}_n\|}-
\frac{\|\vec{x}_{n+1}-\vec{x}_0\|}{R\|\vec{x}-\vec{x}_{n+1}\|}\right)=\frac{Q_0}{R},\quad \vec{x}\in \partial B(\vec{x}_0,R).
\end{equation*}
We have also
\begin{equation*}
U\left(\vec{x}\right)=
\frac{q_0}{\|\vec{x}-\vec{y}_0\|}+Q_0\left(\frac{1}{\|\vec{x}-\vec{x}_0\|}-
\frac{\|\vec{y}_{1}-\vec{y}_0\|}{r\|\vec{x}-\vec{y}_{1}\|}\right)+
\end{equation*}
\begin{equation*}
+\sum_{n=1}^{\infty}Q_n\left(\frac{1}{\|\vec{x}-\vec{x}_n\|}-
\frac{\|\vec{y}_{n+1}-\vec{y}_0\|}{r\|\vec{x}-\vec{y}_{n+1}\|}\right)=
\frac{q_0}{r},\quad \vec{x}\in \partial B(\vec{y}_0,r).
\end{equation*}
Consequently, the conditions 2,3 satisfy too.

\section{\bf The density of the equilibrium distribution}

The density of equilibrium distribution on $\partial B(\vec{x}_0,R)$ denote by
\begin{equation*}
\rho_R(\vec{x}),\quad\vec{x}\in \partial B(\vec{x}_0,R)
\end{equation*}
and the density of equilibrium distribution on $\partial B(\vec{y}_0,r)$ denote by
\begin{equation*}
\rho_r(\vec{x}),\quad\vec{x}\in \partial B(\vec{y}_0,r)
\end{equation*}
For $\rho_R(\vec{x}),\quad\vec{x}\in \partial B(\vec{x}_0,R)$ we have 
\begin{equation*}
\rho_R\left(\vec{x}\right)=\frac{1}{4 \pi R}\sum_{n=0}^{\infty}Q_n
\frac{R^2-\|\vec{x}_0-\vec{x}_n\|^2}{\|\vec{x}-\vec{x}_n\|^3}.
\end{equation*}
similarly for
$\rho_r(\vec{x}),\quad\vec{x}\in \partial B(\vec{y}_0,r)$ 
we have
\begin{equation*}
\rho_r(\vec{x})=\frac{1}{4 \pi r}\sum_{n=0}^{\infty}
q_n\frac{r^2-\|\vec{y}_0-\vec{y}_n\|^2}{\|\vec{x}-\vec{y}_n\|^{3}}.
\end{equation*}
Let us note that the potential function for the equilibrium distribution equals
\begin{equation*}
U(\vec{x})=\frac{Q_0}{R},\quad \vec{x}\in\partial B(\vec{x}_0,R)
\end{equation*}
\begin{equation*}
U(\vec{x})=\frac{q_0}{r},\quad \vec{x}\in \partial B(\vec{y}_0,r).
\end{equation*}
If the point $\vec x$ is outside of the conductor $E$, i.e. 
$\vec{x}\notin E$ then we have:
\begin{equation*}
U(\vec{x})=\frac{1}{4 \pi R^2}\int_{\partial B(\vec{x}_0,R)}
\frac{\rho_R(\vec{y})}{\|\vec{x}-\vec{y}\|}ds
+\frac{1}{4 \pi r^2}\int_{\partial B(\vec{y}_0,r)}
\frac{\rho_r(\vec{y})}{\|\vec{x}-\vec{y}\|}ds.
\end{equation*}

\section{\bf The auxilary results}

Let $\vec{x}_n\in B(\vec{x}_0,R)$ and $\vec{y}_m \in B(\vec{y}_0,r)$. 
Since 
\begin{equation*}
\frac{\vec{x}-\vec{y}_m}{\|\vec{x}-\vec{y}_m\|^3},\quad \vec{x}\in B(\vec{x}_0,R)
\end{equation*}
is harmonic function, so
\begin{equation*}
\frac{1}{4\pi R}\int_{\partial B(\vec{x}_0,R)}
\frac{R^2-\|\vec{x}_0-\vec{x}_n\|^2}{\|\vec{x}-\vec{x}_n\|^{3}}
\left(\frac{\vec{x}-\vec{y}_m}{\|\vec{x}-\vec{y}_m\|^3}\right)ds=
\frac{\vec{x}_n-\vec{y}_m}{\|\vec{x}_n-\vec{y}_m\|^3}.
\end{equation*}
Now let us consider the case if bothe points are inside of the ball $B(\vec{x}_0,R)$.
Let $\vec{x}_n,\,\,\vec{x}_m \in B(\vec{x}_0,R)$. Remember that if $\vec{y}_{m-1}$ is simmetric to $\vec{x}_m$ with respect the sphere $\partial B(\vec{x}_0,R)$ then 
\begin{equation*}
R\|\vec{x}-\vec{x}_m\|=\|\vec{x}_0-\vec{y}_{m-1}\|\|\vec{x}-\vec{y}_{m-1}\|,
\quad \vec{x}\in \partial B(\vec{x}_0,R).
\end{equation*}
Since
\begin{equation*}
\frac{1}{\|\vec{y}_{m-1}-\vec{x}\|},\quad \vec{x}\in B(\vec{x}_0,R)
\end{equation*}
is harmonic function, so
\begin{equation*}
\frac{1}{4\pi R}\int_{\partial B(\vec{x}_0,R)}
\frac{R^2-\|\vec{x}_0-\vec{x}_n\|^2}{\|\vec{x}-\vec{x}_n\|^{3}}
\frac{1}{\|\vec{x}-\vec{x}_m\|}ds=
\end{equation*}
\begin{equation*}
=\frac{R}{\|\vec{x}_0-\vec{y}_{m-1}\|}\frac{1}{4\pi R}
\left(\int_{\partial B(\vec{x}_0,R)}
\frac{R^2-\|\vec{x}_0-\vec{x}_n\|^2}{\|\vec{x}-\vec{x}_n\|^3}
\frac{1}{\|\vec{x}-\vec{y}_{m-1}\|}ds\right)=
\end{equation*}
\begin{equation*}
=\frac{R}{\|\vec{x}_n-\vec{y}_{m-1}\|\|\vec{x}_0-\vec{y}_{m-1}\|}
=\frac{R}{(\|\vec{x}_0-\vec{y}_{m-1}\|-\|\vec{x}_0-\vec{x}_{n}\|)
\|\vec{x}_0-\vec{y}_{m-1}\|}=
\end{equation*}
\begin{equation*}
=\frac{\|\vec{x}_0-\vec{x}_{n}\|\|\vec{x}_0-\vec{x}_{m}\|}
{R(R^2-\|\vec{x}_0-\vec{x}_{n}\|\|\vec{x}_0-\vec{x}_{m}\|)}.
\end{equation*}
Applying the operator $\nabla_{\vec{x}_m}$ we obtain:
\begin{equation*}
-\frac{1}{4\pi R}\int_{\partial B(\vec{x}_0,R)}
\frac{R^2-\|\vec{x}_0-\vec{x}_n\|^2}{\|\vec{x}-\vec{x}_n\|^3}
\frac{\vec{x}_m-\vec{x}}{\|\vec{x}_m-\vec{x}\|^3}ds=
\end{equation*}
\begin{equation*}
=\frac{\vec{y}_0-\vec{x}_0}{\|\vec{y}_0-\vec{x}_0\|}
\frac{R\|\vec{x}_n-\vec{x}_0\|}{(R^2-\|\vec{x}_0-\vec{x}_m\|\|\vec{x}_0-\vec{x}_n\|)^2}.
\end{equation*}

\section{\bf The force of interaction}

	The force acting on the sphere $\partial B(\vec{x}_0,R)$ equals
\begin{equation*}
\vec{F}_R=-\frac{1}{2}\int_{\partial B(\vec{x}_0,R)}
\nabla U(\vec{x})\rho_R(\vec{x})ds=
\end{equation*}
\begin{equation*}
=\frac{1}{8\pi R}\sum_{n=0}^{\infty}\sum_{m=0}^{\infty}Q_nQ_m
\int_{\partial B(\vec{x}_0,R)}
\frac{\vec{x}_m-\vec{x}}{\|\vec{x}-\vec{x}_m\|^3}
\frac{R^2-\|\vec{x}_0-\vec{x}_n\|^2}{\|\vec{x}-\vec{x}_n\|^3}ds+
\end{equation*}
\begin{equation*}
+\frac{1}{8\pi R}\sum_{m=0}^{\infty}\sum_{n=0}^{\infty}q_m Q_n\int_{\partial B(\vec{x}_0,R)}\frac{\vec{y}_m-\vec{x}}{\|\vec{x}-\vec{y}_m\|^3}
\frac{R^2-\|\vec{x}_0-\vec{x}_n\|^2}{\|\vec{x}-\vec{x}_n\|^3}ds=
\end{equation*}
\begin{equation*}
=F_R\frac{\vec{x}_0-\vec{y}_0}{\|\vec{x}_0-\vec{y}_0\|},
\end{equation*}
where
\begin{equation*}
F_R=\sum_{n=0}^{\infty}\sum_{m=0}^{\infty}
\frac{Q_n Q_mR\|\vec{x}_n-\vec{x}_0\|}{2(R^2-\|\vec{x}_0-\vec{x}_m\|
\|\vec{x}_0-\vec{x}_n\|)^2}-
\sum_{n=0}^{\infty}\sum_{m=0}^{\infty}
\frac{q_nQ_m}{2\|\vec{x}_m-\vec{y}_n\|^2}=
\end{equation*}
\begin{equation*}
=\sum_{n=0}^{\infty}\sum_{m=0}^{\infty}
\frac{Q_nQ_mRu(n)}{2(R^2-u(m)u(n))^2}-
\sum_{n=0}^{\infty}\sum_{m=0}^{\infty}
\frac{q_nQ_m}{2(d-u(m)-v(n))^2}.
\end{equation*}
Similarly, The force acting on the ball $B(\vec{y}_0,r)$ equals
\begin{equation*}
\vec{F}_r=-\frac{1}{2}\int_{\partial B(\vec{y}_0,r}
\nabla U(\vec{x})\rho_r(\vec{x})ds=F_r
\frac{\vec{y}_0-\vec{x}_0}{\|\vec{y}_0-\vec{x}_0\|},
\end{equation*}
where
\begin{equation*}
F_r=\sum_{n=0}^{\infty}\sum_{m=0}^{\infty}
\frac{rq_nq_m\|\vec{y}_0-\vec{y}_n\|}
{2(r^2-\|\vec{y}_0-\vec{y}_n\|\|\vec{y}_0-\vec{y}_m\|)^2}-
\sum_{n=0}^{\infty}\sum_{m=0}^{\infty}
\frac{q_nQ_m}{2\|\vec{y}_m-\vec{x}_n\|^2}=
\end{equation*}
\begin{equation*}
=-\sum_{n=0}^{\infty}\sum_{m=0}^{\infty}\frac{Q_nq_m}
{2(d-u(n)-v(m))^2}+\sum_{n=0}^{\infty}\sum_{m=0}^{\infty}
\frac{q_n q_m r v(m)}{2(r^2-v(n)v(m))^2}.
\end{equation*}
In numerical calculations it is useful to take into account the following inequality
\begin{equation*}
\left\|\vec{F_R}-\sum_{n=0}^{N}\sum_{m=0}^{N}
\frac{Q_nQ_mRu(n)}{2(R^2-u(m)u(n))^2}+
\sum_{n=0}^{N}\sum_{m=0}^{N}\frac{q_nQ_m}{2(d-u(m)-v(n))^2}\right\|\leq
\end{equation*}
\begin{equation*}
\leq \sum_{n=N+1}^{\infty}\sum_{m=0}^{\infty}\frac{|Q_n|| q_m|}{(d-u(n)-v(m))^2}+\sum_{n=0}^{\infty}\sum_{m=N+1}^{\infty}
\frac{|Q_n||Q_m| u(n)}{2(r^2-u(n)u(m))^2}\leq
\end{equation*}
\begin{equation*}
\leq A\left(\frac{Rr}{(d-R)(d-r)}\right)^{\frac{N}{2}}.
\end{equation*}
So the approximation converges as geometric progression.
Similar results we have for $\vec{F}_r$.

\section{Conclusion}

In some values of parameters, one of the ball vibrates. Note that in considered model the forces that keep the charges on the boundaries of the balls are not explicitly presented. So the system is not close and the Newton's law can not satisfied.

\vskip 1cm

\section{Appendix.}

TwoBalls

for n = 2:N+1;

    u(n) = r*r/(d - v(n-1));

    v(n) = R*R/(d - u(n-1));

end

Q1 = ones(1, N+1);

q1 = ones(1, N+1);

for n = 2:N+1;

    Q1(n) = -q1(n-1)*u(n)/R;

    q1(n) = -Q1(n-1)*v(n)/r;

end

M = [sum(Q1(2:2:N)), sum(Q1(1:2:N));

      sum(q1(2:2:N)), sum(q1(1:2:N));];

A = M/Q;

QQ = A(1, 1).*Q1;

qq = A(2, 1).*q1;

[U, V] = ndgrid(u, v);

[n1, m1]=ndgrid(1:N+1,1:N+1);

FR11 = (QQ'*qq)./(2*(d - V - U).*(d - V - U));

FR12 = (QQ'*qq).*U.*R./(2*(R*R - U.*U').*(R*R - U.*U'));

Fr21 = (qq'*QQ)./(2*(d - V - U).*(d - V - U));

Fr22 = (qq'*qq).*V.*r./(2*(r*r - V'.*V).*(r*r - V'.*V)));

FR = sum(FR11(:))-sum(FR12(:));  %force direction x0-y0

Fr = sum(Fr21(:))-sum(Fr22(:));  %force direction y0-x0

testTwoBalls

c =0.01:0.01:5;

FR = [];

Fr = [];

for k = 1:numel(c);

    [FR(k), Fr(k)] = TwoBalls(c(k));

end

for k=1:numel(c);

    O(k)=0;

end

figure(1)
k=30:1:numel(c);
plot(k,FR(k),'r',k,Fr(k),'b',k,O(k),'y')

figure(2)
k=1:1:30;
plot(k,FR(k),'r',k,O(k),'y')

figure(3)
k=1:1:30;
plot(k,Fr(k),'b',k,O(k),'y')

\vskip 1 cm

\begin{figure}[h]	
\centering
	\includegraphics [width=0.7\textwidth]{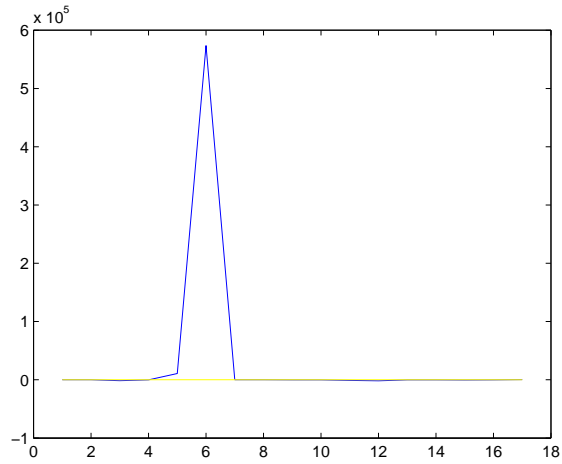}
	\caption{Let R=1 be the radius of the big ball and r=0.8 be radius of the small ball. Let d=R+r+ck be the distance of the ball's centers, 
where $c=0.001$, and $k=1:17$. In the picture you see $F_r$.}
\end{figure}

\begin{figure}[h]	
\centering
	\includegraphics [width=0.7\textwidth]{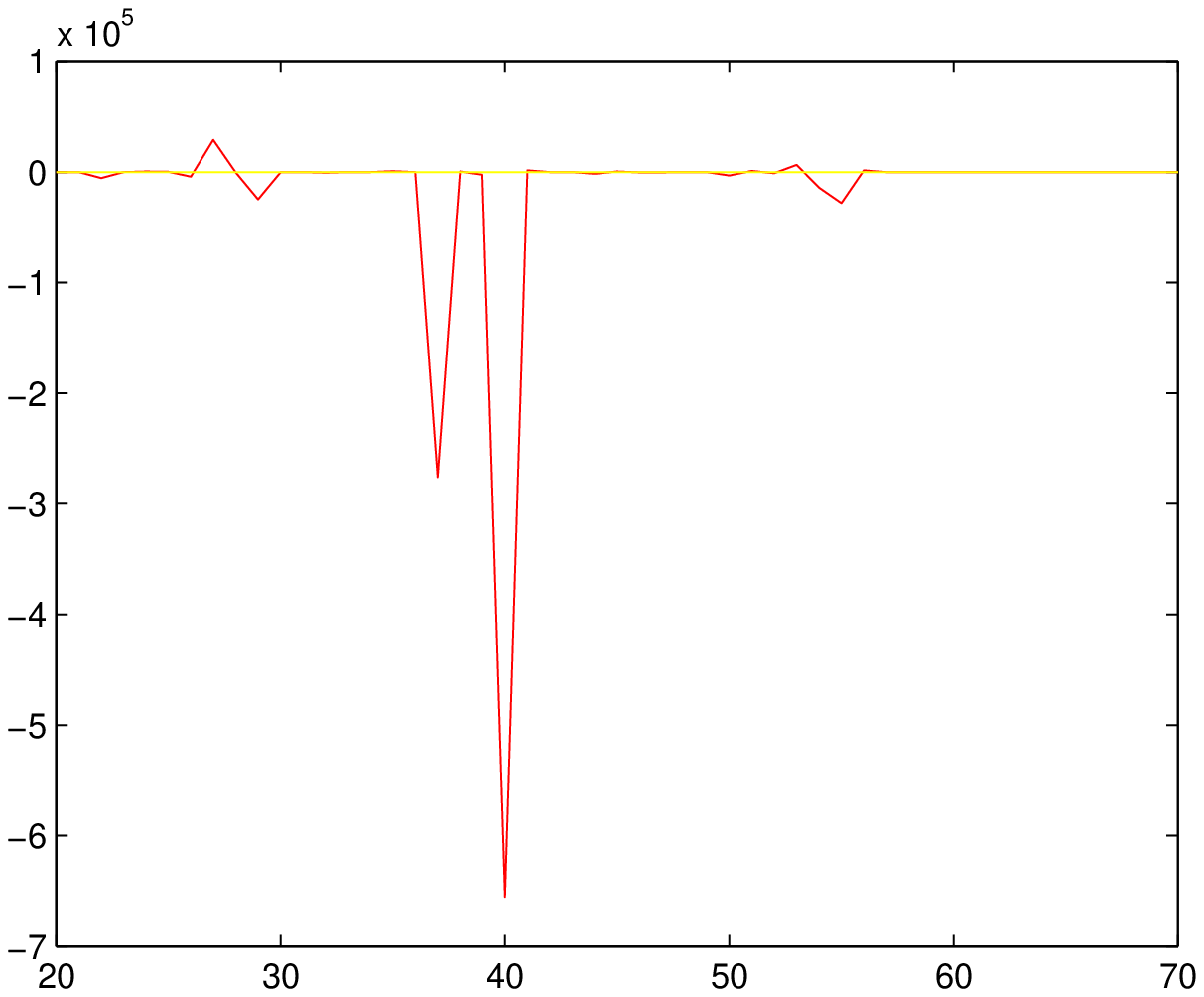}
	\caption{Let R=1 be the radius of the big ball and r=0.8 be radius of the small ball. Let d=R+r+ck be the distance of the ball's centers, 
where $c=0.001$, and $k=20:70$. In the picture you see $F_r$.}
\end{figure}

\begin{figure}[h]	
\centering
	\includegraphics [width=0.7\textwidth]{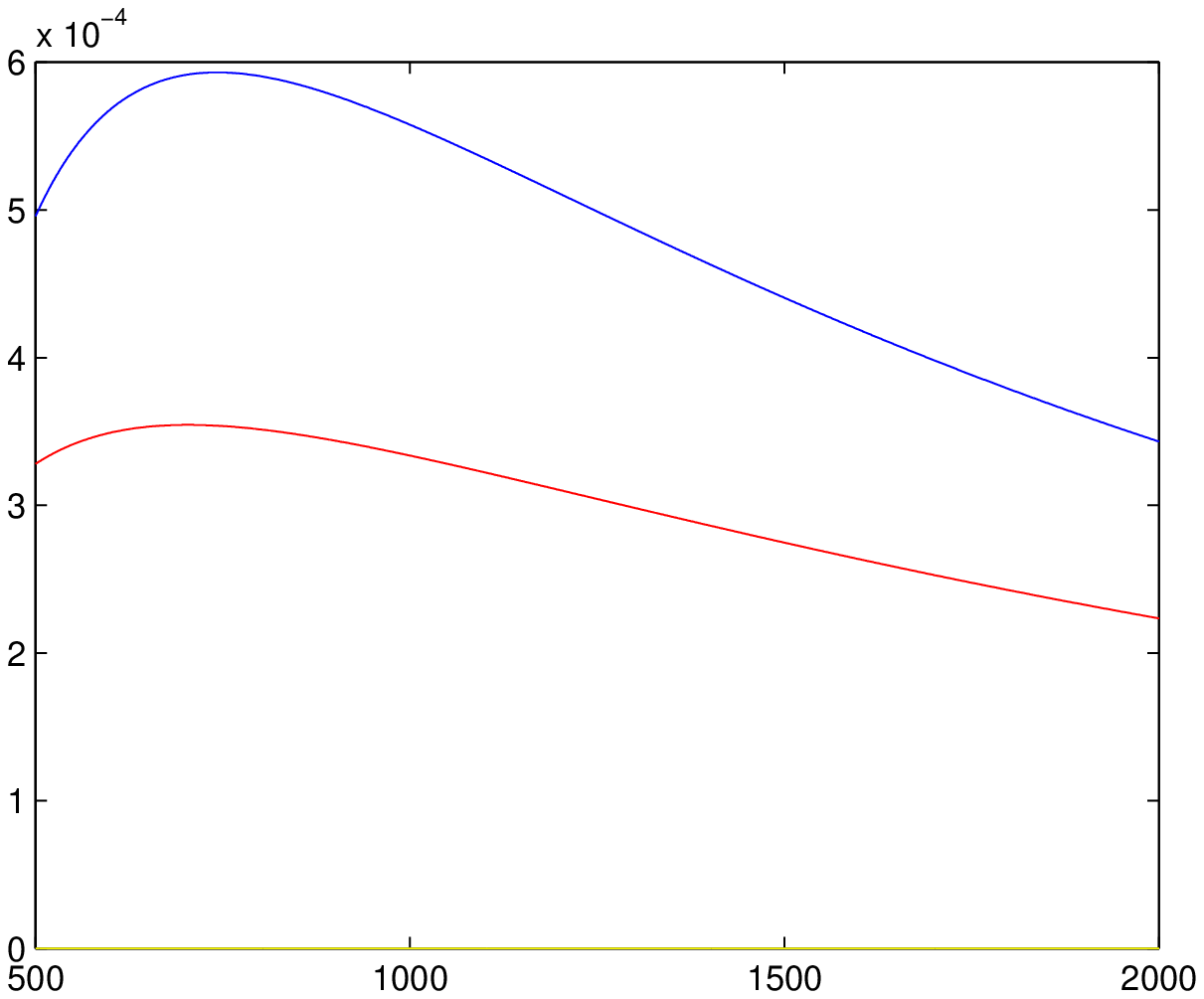}
	\caption{Let R=1 be the radius of the big ball and r=0.8 be radius of the small ball. Let d=R+r+ck be the distance of the ball's centers, 
where $c=0.001$, and $k=500:2000$. In the picture you see $F_r$ and $F_R$ for big distences.}
\end{figure}
\end{document}